\documentclass[aps,twocolumn,showpacs,amsmath,amssymb]{revtex4-1}
\usepackage{graphicx}
\usepackage{dcolumn}
\usepackage{bm}
\usepackage{hyperref}

\begin{document}

\title{Robustness of genuine tripartite entanglement under collective dephasing}
\author{Mazhar Ali}
\affiliation{Department of Electrical Engineering, Faculty of Engineering, Islamic University in Madinah, 107 Madinah, Kingdom of Saudi Arabia}

\begin{abstract}
We study robustness of genuine multipartite entanglement for a system of three qubits under collective dephasing. 
Using a computable entanglement monotone for multipartite systems, we found that almost every state is quite robust under this
type of decoherence. We analyze random states and weighted graph states at infinity and found all of them to be 
genuinely entangled. 
\end{abstract}


\maketitle

Quantum entanglement is essential for several practical applications in the emerging field of quantum Information. This feature 
inspire us to develop a theory of entanglement and several methods to partially avoid the unalterable process of 
decoherence \cite{Horodecki-RMP-2009, gtreview}. As we desire to generate and manipulate entanglement in experiments, 
therefore it is essential to study the effects of various environments on entanglement \cite{Aolita-review}.
This problem has received considerable attention for bipartite and multipartite systems
\cite{Yu-work,lifetime,Aolita-PRL100-2008,bipartitedec,Band-PRA72-2005,lowerbounds,Lastra-PRA75-2007, Guehne-PRA78-2008,Lopez-PRL101-2008, 
Ali-work, HI-JPA43-2010, Weinstein-PRA85-2012, Wu-PRA89-2014}.
Few studies considered the life time of entanglement under decoherence \cite{Yu-work,lifetime}, however life time may not characterize 
the decoherence process appropriately. The life time of entanglement may be long, but the actual amount of entanglement might be very small 
after a short time, making the decohered state practically useless for information processing tasks \cite{Aolita-PRL100-2008}. 
Some authors have studied only bipartite aspects of entanglement of multipartite states \cite{Band-PRA72-2005}, nevertheless these studies can capture 
only partial information because genuine entanglement is distinct in nature than the entanglement among various partitions.
Unless there is an adequate theory of multipartite entanglement, one can talk about lower bounds of entanglement instead of its exact 
value \cite{lowerbounds}. The precise expression of multipartite entanglement was only computed for particular type of decoherence and definite 
quantum states \cite{Guehne-PRA78-2008}. In addition, in order to better judge the dynamical behaviour of a state, we need to compare 
its dynamics with dynamics of random states. In our previous studies, we focussed on the dynamics of multipartite states by investigating a 
computable measure of genuine multipartite entanglement for several decoherence models 
\cite{Ali-work-2014}.

In this Letter we extend our studies to examine the robustness of three-qubit states as well as random states under 
collective dephasing. Recent progress in the theory of multipartite entanglement has enabled us to study decoherence effects on actual multipartite 
entanglement and not on entanglement among bipartitions. In particular, the ability to compute genuine negativity for multipartite systems 
has eased this task \cite{Bastian-PRL106-2011, Hofmann-JPA47-2014}. 
We find that due to collective damping effects, decoherence-free-subspace exist in this system. This space is spanned by the W-type states. 
Hence the W-type states are most robust being uneffected by decoherence process \cite{Doll-IJQI6-2008}. 
Although the entanglement of GHZ states is degraded by decoherence process, nevertheless, they are also quite robust and loose their genuine 
entanglement only at infinity. Interestingly, almost every pure random state is robust under collective dephasing. 

We consider the three qubits to be spins for example and impose a stochastic magnetic field $B(t)$ acting on three qubits together. 
In the interaction picture, the Hamiltonian of the qubits and classical noisy field can be written as (with $\hbar = 1$)
\begin{equation}
H(t) = - \frac{1}{2} \mu \, B(t) (\sigma_z^A + \sigma_z^B + \sigma_z^C), 
\label{Eq:Ham}
\end{equation}
where $\mu$ is the gyromagnetic ratio and $\sigma_z^{A,B,C}$ are the standard Pauli matrices in the computational basis 
$\{ |0\rangle, \, |1\rangle\}$. The stochastic magnetic field refer to statistically independent classical Markov process satisfying 
the conditions
\begin{eqnarray} 
\langle B(t) \, B(t')\rangle &=& \frac{\Gamma}{\mu^2} \, \delta(t-t') \,, \nonumber \\ \langle B(t)\rangle &=& 0 \, ,
\label{Eq:tavf}
\end{eqnarray}
where $\langle \cdots \rangle$ is ensemble time average and $\Gamma$ denote the phase damping rate for collective dephasing.

We choose the standard computational basis $ \{ \, |000\rangle$, $|001\rangle$,$\ldots$, $|111\rangle \, \}$. The time-dependent 
density matrix for three-qubits can be obtained by taking 
ensemble average over the noise field, i.\,e., $\rho(t) = \langle\rho_{st}(t)\rangle$, 
where $\rho_{st}(t) = U(t) \rho(0) U^\dagger(t)$ and $U(t) = \exp[-\mathrm{i} \int_0^t \, dt' \, H(t')]$. 
We assume that there are no initial correlations between the principal system and stochastic field, that is, 
$\rho(0) = \rho_S \otimes \rho_R$. There are various ways to compute the time evolved density matrix for 
three qubits, however we prefer the master equation approach to solve the system. 

We consider a system "S" interacting with a reservoir "R". The combined density operator is given as $\rho_{SR}$. The reduced 
density matrix $\rho_S$ for system alone can be obtained by taking trace over reservoir degrees of freedom, that is, 
$\rho_S = {\rm Tr}_R (\rho_{SR})$. In the interaction picture, the equation of motion is given as (with $\hbar = 1$) 
\begin{eqnarray}
i \dot{\rho}_{SR} = \big[ H, \, \rho_{SR}(t)\big]. \label{A1} 
\end{eqnarray}
As the interaction between system and reservoir is weak, therefore we look solution of the form 
$\rho_{SR}(t) = \rho_S(t) \otimes \rho_R(t_{\rm i}) + \rho_c(t)$, where for consistency we demand that ${\rm Tr}_R(\rho_c(t)) = 0$. 
Substituting this form and Hamiltonian (\ref{Eq:Ham}), retaining terms upto order $H^2$, and after doing some simplification, we get
\begin{eqnarray}
\frac{d \rho(t)}{dt} =& - \frac{\Gamma}{8} \bigg\{ (\sigma_z^A + \sigma_z^B + \sigma_z^C)(\sigma_z^A 
+ \sigma_z^B + \sigma_z^C) \rho(t) \nonumber \\& 
- 2 \, (\sigma_z^A + \sigma_z^B + \sigma_z^C)\,\rho(t) \, (\sigma_z^A + \sigma_z^B + \sigma_z^C) \nonumber \\& 
+ \rho(t) (\sigma_z^A + \sigma_z^B + \sigma_z^C) (\sigma_z^A + \sigma_z^B + \sigma_z^C) \, \bigg\} . \label{A6}
\end{eqnarray}
We have used the relations (\ref{Eq:tavf}) to obtain the above equation. It is now straight forward to obtain the most general solution. 
Due to collective dephasing, there appear the decoherence free subspaces (DFS) for W type states \cite{Doll-IJQI6-2008}. This DFS is spanned by 
basis $\{ \, |001\rangle$, $|010\rangle$, $|100\rangle$, $|011\rangle$, $|101\rangle$, $|110\rangle \}$. Hence the W type 
states are most robust states under this type of decoherence model.

To present the concept of genuine entanglement, we take a system of three parties $A$, $B$, and $C$. 
A state is defined as biseparable if it can be written as a mixture of states which are separable with respect
to different bipartitions, that is
\begin{eqnarray}
\rho^{bs} = p_1 \, \rho_{A|BC}^{sep} + p_2 \, \rho_{B|AC}^{sep} + p_3 \, \rho_{C|AB}^{sep}\,.
\end{eqnarray}
A state is genuinely multipartite entangled if it is not biseparable. This definition can readily be extended to $N$-partite 
systems.

Genuine entanglement can be detected and characterized via the technique of positive partial transpose mixtures 
(PPT mixtures) \cite{Bastian-PRL106-2011}. A bipartite state 
$\rho = \sum_{ijkl} \, \rho_{ij,kl} \, |i\rangle\langle j| \otimes |k\rangle\langle l|$ is PPT if its
partially transposed matrix
$\rho^{T_A} = \sum_{ijkl} \, \rho_{ji,kl} \, |i\rangle\langle j| \otimes |k\rangle\langle l|$ is positive semidefinite. As separable states 
are always PPT \cite{peresPPT} and therefore the set of separable states with respect to some partition is contained in a larger set of 
states which has a positive partial transpose for that bipartition. The states which are PPT with respect to fixed bipartition 
may be called $\rho_{A|BC}^{PPT}$, $\rho_{B|AC}^{PPT}$,
and $\rho_{C|AB}^{PPT}$. We ask whether a state can be written as a mixing of PPT states 
\begin{eqnarray}
\rho^{PPTmix} = p_1 \, \rho_{A|BC}^{PPT} + p_2 \, \rho_{B|AC}^{PPT} + p_3 \, \rho_{C|AB}^{PPT}\,,
\end{eqnarray}
which is called a PPT mixture.
Because any biseparable state is a PPT mixture, any state which is not a PPT mixture must be genuinely
entangled. The main benefit of employing PPT mixtures instead of biseparable states is that
PPT mixtures can be fully characterized by semidefinite programming (SDP) \cite{sdp}. In addition, the set of
PPT mixtures serves a very good approximation to the set of biseparable states and delivers the best known separability criteria for
many cases, nevertheless there are multipartite entangled states which are PPT mixtures \cite{Bastian-PRL106-2011}. 
A state is a PPT mixture if and only if 
\begin{eqnarray}
 \min {\rm Tr} (\mathcal{W} \rho)
\end{eqnarray}
has a positive solution under the constraint that for all bipartitions $M|\bar{M}$
\begin{eqnarray}
 \mathcal{W} = P_M + Q_M^{T_M},
 \, \mbox{ with }
 0 \leq P_M\,\leq I ,
 0 \leq  Q_M  \leq I.
\end{eqnarray}
The constraints reflect that $\mathcal{W}$ is a decomposable entanglement
witness for any bipartition. If this minimum is negative then $\rho$ is not a PPT mixture and hence genuinely 
entangled \cite{Bastian-PRL106-2011}. Since this is a semidefinite program, the minimum can be efficiently computed and the 
optimality of the solution can be certified \cite{sdp}. We use the programs YALMIP and SDPT3 \cite{yalmip} to solve SDP. 
We also use implementation which is freely available \cite{PPTmix}. For more explanations, see Ref.\cite{Ali-work-2014}.

The absolute value of the minimization was shown to be an entanglement monotone \cite{Bastian-PRL106-2011}. 
This monotone has been called genuine negativity \cite{Hofmann-JPA47-2014}. In this work we use this measure by denoting it $E(\rho)$. 
For bipartite systems, it is equivalent to {\it negativity} \cite{Vidal-PRA65-2002} and for a system of qubits, it is bounded by 
$E(\rho) \leq 1/2$ \cite{bastiangraph}.

Two inequivalent genuinely entangled states for three qubits are 
the GHZ states $|GHZ_1 \rangle = 1/\sqrt{2}(|000\rangle + |111\rangle)$ and the W states 
$|W\rangle = 1/\sqrt{3}(|001\rangle + |010\rangle + |100\rangle)$. 
Entanglement monotone for GHZ state has a value of $E(|GHZ\rangle\langle GHZ|) = 1/2$, whereas for the W state, its 
value is $E(|W_3\rangle\langle W_3|) \approx 0.443$. As the W state lives in DFS, therefore it is already robust, hence we only focus 
on GHZ states. Another locally equivalent GHZ state is $ |GHZ_2 \rangle = 1/\sqrt{2}(|001\rangle + |110\rangle)$.
In order to make meaningful comments on robustness of these states, we compare the dynamics with that of 
random pure states and weighted graph states. The method to generate these random states is described in detail in Ref.\cite{Ali-work-2014}.

We argued before that the life-time of entanglement may not lead to conclusive results. In addition, starting with initial states 
having different amount of entanglement, the lifetime of entanglement may not tell the relative decay. Moreover, the comparison of the values of
$E[\rho_i(t)]$ for different $t$ and initial states $i$ is not useful. To overcome these problems, we may consider the logarithmic derivative 
\begin{eqnarray}
\eta(t) = \frac{d\,(\ln[E(t)])}{dt} = \frac{d/dt \,[E(t)]}{E(t)} \, , 
\end{eqnarray}
where $E(t)$ is the entanglement monotone \cite{Guehne-PRA78-2008}. 
This quantity describes the relative decay of entanglement, and it is now possible to compare states 
with different initial entanglement. We note that for a state where the entanglement just decays exponentially, $\eta(t)$
is constant and the inverse of the half-life. We stress that as in our previous study \cite{Ali-work-2014}, 
we use logarithmic derivative of genuine entanglement to claim the robustness of all states. 
The term "robustness of entanglement" was used in Refs.~\cite{Vidal-PRA59-1999,Borras-PRA79-2009,Novotny-PRL107-2011}, but 
in \cite{Vidal-PRA59-1999} it was used as a kind of quantification of entanglement.
In \cite{Borras-PRA79-2009}, the authors have identified most robust states under local decoherence used the definition introduced 
before \cite{Band-PRA72-2005}, whereas, the authors of Ref.\cite{Novotny-PRL107-2011} considered asymptotic long-time dynamics of initial 
states and identified two different classes of states, one class is fragile even there remains some coherence in the system and the second 
class as most robust states which become disentangled only when decoherence is perfect. However, our approach is different than these 
studies.

\begin{figure}[t!]
\scalebox{1.97}{\includegraphics[width=1.87in]{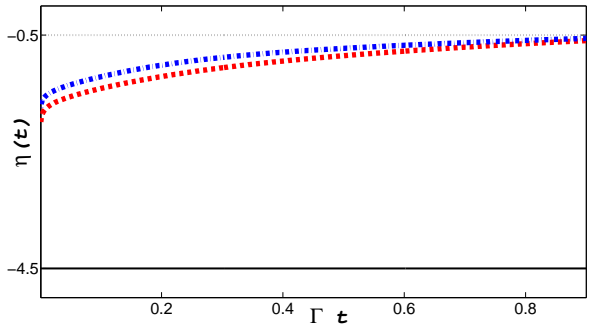}}
\centering
\caption{Logarithmic derivative is plotted against parameter $\Gamma t$ GHZ states, mean values for weighted graph states and random pure 
states.}
\label{Fig:R3QbGHZ1}
\end{figure}
Figure \ref{Fig:R3QbGHZ1} shows logarithmic derivative plotted 
against parameter $\Gamma t$. The constant dotted line at $-0.5$ is for $|GHZ_2\rangle$ state whereas constant solid line at $-4.5$ is for 
$|GHZ_1\rangle$ state. This means that both states remain genuinely entangled till infinity and decay only exponentially. 
This claim can also be verified from the fact that as all zero elements in the density matrix remain zero, therefore
GHZ states is genuinely entangled if $|\rho_{27}| = e^{- \Gamma t/2}/2 > 0$ for $|GHZ_2\rangle$ state and 
$|\rho_{18}| = e^{- 9/2 \Gamma t}/2  > 0$ for $|GHZ_1\rangle$ state \cite{Otfried-NJP12-2010}, which is obviously the case here. 
The thick dashed line is the mean value of logarithmic derivative for $100$ random pure states whereas the thick dashed-dotted line 
is the corresponding mean value for $100$ weighted graph states. 
We observe that these two mean values converge to the value of $|GHZ_2\rangle$ state. 
This means that almost every random state is quite robust. This interesting feature is due to presence of decoherence-free-subspaces (DFS).

\begin{figure}[t!]
\scalebox{1.97}{\includegraphics[width=1.87in]{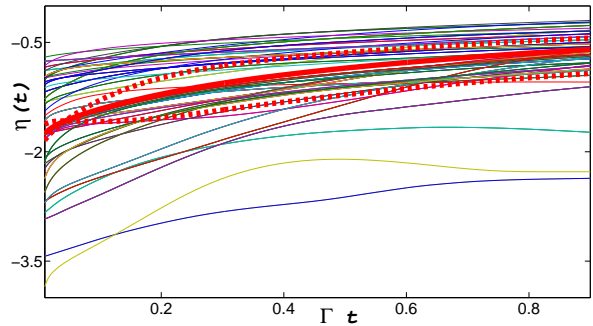}}
\centering
\caption{Logarithmic derivative for random pure states plotted against parameter $\Gamma t$. See text for details.}
\label{Fig:R3QbCD2}
\end{figure}

In Figure \ref{Fig:R3QbCD2}, we plot logarithmic derivative $\eta(t)$ for $100$ random pure states. We observe that every random state is 
robust under collective damping. We also plot the mean value of $\eta(t)$ (thick solid line). From this data we also obtain an 
error estimate to indicate the reliability of measure. This can, for instance, be defined as a confidence interval
\begin{eqnarray}
 CI = \mu \, \pm \, \sqrt{\delta} \, ,
 \label{interval}
\end{eqnarray}
where $\mu$ stands for mean value and $\delta$ for variance of quantity being measured. Note, however, that this is not a confidence
interval in the mathematical sense. The thick dashed lines are the confidence intervals for the mean value.

\begin{figure}[t!]
\scalebox{1.97}{\includegraphics[width=1.87in]{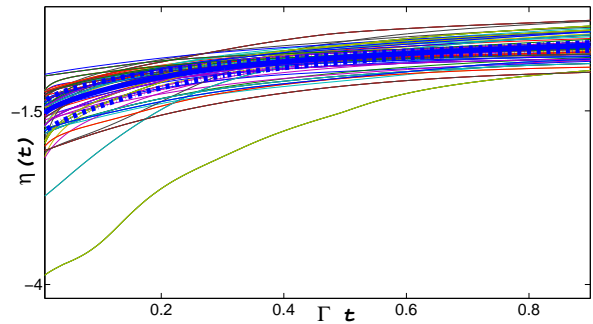}}
\centering
\caption{Logarithmic derivative for weighted graph states plotted against parameter $\Gamma t$. See text for details.}
\label{Fig:R3QbCD3}
\end{figure}
Figure \ref{Fig:R3QbCD3} shows the logarithmic derivative for $100$ weighted graph states plotted against $\Gamma t$. As in the case of random 
pure states, every state here tends towards the value of $-0.5$. The thick solid line is the mean value of $\eta(t)$ whereas 
dashed lines are the corresponding confidence intervals. 

\begin{figure}[t!]
\scalebox{2.0}{\includegraphics[width=1.9in]{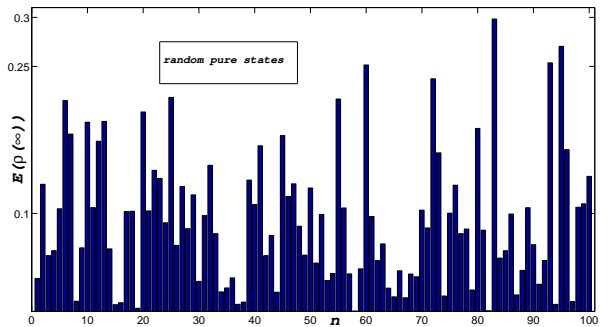}}
\centering
\caption{Entanglement monotone for asymptotic random states plotted against parameter $n$.}
\label{Fig:EM3QbRSAYP1}
\end{figure}
Figure \ref{Fig:EM3QbRSAYP1} is the bar plot of entanglement monotone for $n = 100$ random pure states at $t = \infty$. It can be seen 
that each state has strictly positive value of entanglement monotone. Hence all random pure states remain genuinely entangled even at infinity.

\begin{figure}[t!]
\scalebox{2.0}{\includegraphics[width=1.9in]{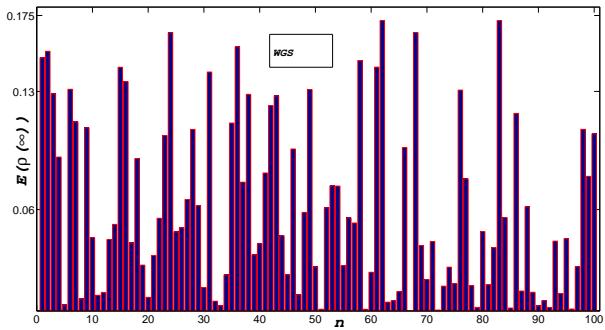}}
\centering
\caption{Entanglement monotone for asymptotic weighted graph states plotted against parameter $n$.}
\label{Fig:EM3QbWGSAYP1}
\end{figure}
Figure \ref{Fig:EM3QbWGSAYP1} depicts the bar plot of entanglement monotone for $n = 100$ weighted graph states at $t = \infty$. Once again all 
states remain genuinely entangled forever, however, the amount of entanglement monotone is lower as compared with random pure states.

We analyze the entanglement properties of asymptotic random states where the complete decoherence has occured and $\gamma = e^{-\Gamma t/2} = 0$. 
The non-vanishing density matrix elements at $t = \infty $ are 
$ \big\{ \rho_{11}$, $\rho_{22}$, $\rho_{23}$, $\rho_{25}$, $\rho_{32}$, $\rho_{33}$, $\rho_{35}$, $\rho_{44}$, $\rho_{46}$, $\rho_{47}$, 
$\rho_{52}$, $\rho_{53}$, $\rho_{55}$, $\rho_{64}$, $\rho_{66}$, $\rho_{67}$, $\rho_{74}$, $\rho_{76}$, $\rho_{77}$, $\rho_{88} \big\}$. 

We have studied the robustness of genuine multipartite entanglement under collective dephasing. Due to appearence of 
decoherence free subspaces, our findings show that almost every quantum state in this type of dynamics is robust. 
As the $W$ type states live in this DFS, therefore they are effectly immune to decoherence and hence most robust states. 
Interestingly,  We have compared the dynamics of specific states with dynamics of 
random pure states and weighted graph states. Again, we found that every single state thus generated is robust and remain 
genuinely entangled throughout the dynamical process. This situation is in contrast with two qubits case where there are always some fragile quantum 
states under collective dephasing. We believe that our findings would be motivational for an experimental verification.

The author is grateful to referees for their positive comments and suggestions.


\begin{thebibliography}{99}

\bibitem{Horodecki-RMP-2009} Horodecki R, Horodecki P, Horodecki M and Horodecki K 2009 Rev. Mod. Phys. {\bf 81} 865

\bibitem{gtreview} G\"uhne O and T\'oth G 2009 Phys. Rep. {\bf 474} 1

\bibitem{Aolita-review} Aolita L, de Melo F and Davidovich L 2015 Rep. Prog. Phys. {\bf 78} 042001

\bibitem{Yu-work} Yu T and Eberly J H 2002 Phys. Rev. B {\bf 66} 193306; 
Yu T and Eberly J H 2003 Phys. Rev. B {\bf 68} 165322; Yu T and Eberly J H 2004 Phys. Rev. Lett. {\bf 93} 140404;
Eberly J H and Yu T 2007 Science {\bf 316} 555

\bibitem{lifetime} D{\"u}r W and Briegel H J 2004 Phys. Rev. Lett. {\bf 92} 180403;
Hein M, D\"ur W and Briegel H J 2005 Phys. Rev. A {\bf 71} 032350

\bibitem{Aolita-PRL100-2008} Aolita L, Chaves R, Cavalcanti D, Ac\'in A and Davidovich L 2008 Phys. Rev. Lett. {\bf 100} 080501

\bibitem{bipartitedec} Simon C and Kempe J 2002 Phys. Rev. A {\bf 65} 052327;
Borras A {\it et al.} 2009 Phys. Rev. A {\bf 79} 022108;
Cavalcanti D {\it et al.} 2009 Phys. Rev. Lett. {\bf 103} 030502

\bibitem{Band-PRA72-2005} Bandyopadhyay S and Lidar D A 2005 Phys. Rev. A {\bf 72} 042339; 
Chaves R and Davidovich L 2010 Phys. Rev. A {\bf 82} 052308; Aolita L {\it et al.} 2010 Phys. Rev. A {\bf 82} 032317

\bibitem{lowerbounds} Carvalho A R R, Mintert F and Buchleitner A 2004 Phys. Rev. Lett. {\bf 93} 230501

\bibitem{Lastra-PRA75-2007} Lastra F, Romero G, Lopez C E, Fran\c ca Santos M and Retamal J C 2007 Phys. Rev A {\bf 75} 062324

\bibitem{Guehne-PRA78-2008} G\"uhne O, Bodoky F and Blaauboer M 2008 Phys. Rev. A {\bf 78} 060301(R) 

\bibitem{Lopez-PRL101-2008} L\'opez C E, Romero G, Lastra F, Solano E and Retamal J C 2008 Phys. Rev. Lett. {\bf 101} 080503

\bibitem{Ali-work} Rau A R P, Ali M and Alber G 2008 EPL {\bf 82} 40002;
Ali M, Alber G and Rau A R P 2009 J. Phys. B: At. Mol. Opt. Phys. {\bf 42} 025501 
Ali M 2010 J. Phys. B: At. Mol. Opt. Phys. {\bf 43} 045504; Ali M 2010 Phys. Rev. A {\bf 81} 042303;
Ali M 2014 Chin. Phys. B Vol. {\bf 23} No. 9 090306; Ali M and Jiang H 2014 Chin. Phys. Lett. Vol. {\bf 31} No. 11 110301

\bibitem{HI-JPA43-2010} Hamadou-Ibrahim A, Plastino A R and Zander C 2010 J. Phys. A: Math. Gen. {43} 055305;
He Z, Zou J, Shao B and Kong S Y 2010 J. Phys. B: At. Mol. Opt. Phys. {\bf 43} 115503

\bibitem{Weinstein-PRA85-2012} Weinstein Y S {\it et al.} 2012 Phys. Rev. A {\bf 85} 032324; 
Filippov S N, Ryb\'ar T and Ziman M 2012 Phys. Rev. A {\bf 85} 012303; 
Filippov S N, Melnikov A A and Ziman M 2013 Phys. Rev. A {\bf 88} 062328

\bibitem{Wu-PRA89-2014} Yang M-J and Wu S-T 2014 Phys. Rev. A {\bf 89} 022301; Wu S-T 2014 Phys. Rev. A {\bf 89} 034301

\bibitem{Ali-work-2014} Ali M and G\"uhne O 2014 J. Phys. B: At. Mol. Opt. Phys. {\bf 47} 055503; 
Ali M 2014 Phys. Lett. A {\bf 378} 2048; Ali M and Rau A R P 2014 Phys. Rev. A {\bf 90} 042330; 
Ali M 2014 Open. Sys. \& Info. Dyn. Vol. {\bf 21} No. 4 1450008

\bibitem{Bastian-PRL106-2011} Jungnitsch B, Moroder T and G\"uhne O 2011 Phys. Rev. Lett. {\bf 106} 190502;
Novo L, Moroder T and G\"uhne O 2013 Phys. Rev. A {\bf 88} 012305

\bibitem{Hofmann-JPA47-2014} Hofmann M, Moroder T and G\"uhne O 2014 J. Phys. A: Math. Theor. {\bf 47} 155301

\bibitem{Doll-IJQI6-2008} Doll R, Wubs M, Kohler S and H\"anggi P 2008 Int. J. Quant. Info. {\bf 6} 681 

\bibitem{peresPPT} Peres A 1996 Phys. Rev. Lett. {\bf 77} 1413

\bibitem{sdp} Vandenberghe L and Boyd S 1996 SIAM Rev. {\bf 38} 49

\bibitem{yalmip} L\"ofberg J 2004 YALMIP: YALMIP: a toolbox for modeling and
optimization in MATLAB CACSD’04: Proc. Computer Aided Control System Design Conf. (Taipei, Taiwan, 4 September 2004) pp 284–9

\bibitem{PPTmix} {\tt mathworks.com/matlabcentral/fileexchange/30968}

\bibitem{Vidal-PRA65-2002} Vidal G and Werner R F 2002 Phys. Rev. A {\bf 65} 032314

\bibitem{bastiangraph} Jungnitsch B, Moroder T and G\"uhne O 2011 Phys. Rev. A {\bf 84} 032310

\bibitem{Vidal-PRA59-1999} Vidal G and Tarrach R 1999 Phys. Rev. A {\bf 59} 141

\bibitem{Borras-PRA79-2009} Borras A, Majtey A P, Plastino A R, Casas M and Plastino A 2009 Phys. Rev. A {\bf 79} 022108

\bibitem{Novotny-PRL107-2011} Novotn\'y J, Alber G and Jex I 2011 Phys. Rev. Lett. {\bf 107} 090501

\bibitem{Otfried-NJP12-2010} G\"uhne O and Seevinck M 2010 N. J. Phys. {\bf 12} 053002

\end{thebibliography}
\end{document}